# Possible Realization and Protection of Valley-Polarized Quantum Hall Effect in Mn/WS$_2$

Jie Li, Lei Gu and Ruqian Wu*

*Department of Physics and Astronomy, University of California, Irvine, California 92697-4575, USA.*

**ABSTRACT:** By using the first-principles calculations and model analyses, we found that the combination of defected tungsten disulfide monolayer and sparse manganese adsorption may give a KK´ valley spin splitting up to 210 meV. This system also has a tunable magnetic anisotropy energy, a clean band gap, and an appropriate band alignment, with the Fermi level sitting right above the top of valence bands at the K-valleys. Therefore, it can be used for the realization of the valley-polarized anomalous Hall effect and for the exploration of other valley related physics without using optical methods. A protective environment can be formed by covering it with a hexagonal BN layer, without much disturbance to the benign properties of Mn/WS$_2$.

Valley-dependent electronics and optoelectronics have attracted extensive research interest in recent years due to their potential applications in novel quantum information technologies. Like charge and spin, valley is a new degree of freedom of carriers as those from opposite valleys in the Brillouin zone (BZ) counter-propagate along the edges of a two-dimensional material without dissipation.[1,2] Transition-metal dichalcogenides (TMDs) with inversion symmetry breaking and strong spin-orbit coupling (SOC) are the prototypical platforms for studies of the valley-spin physics. Many intriguing phenomena such as Valley Hall effect in monolayer $MoS_2$ transistors,[3] small valley splitting in monolayer $WSe_2$,[4,5] valley-polarized electroluminescence in light-emitting van der Waals heterojunction[6,7] have been reported recently. Furthermore, more possibilities of observing valley-dependent electronics and optoelectronics have been proposed theoretically.[8-12] However, most measurements and manipulations of the valley related phenomena rely on optical excitation, which is inconvenient for operations at a small length scale. For the development of the valley-spin based devices, lifting the valley degeneracy is one of the most important tasks so that carriers in inequivalent valleys can be selectively produced with easier and more localized means such as gating. With an adequately large valley splitting, the valley Hall effect can manifest as we move the Fermi level to touch the valence band maximum (VBM) or the conduction band minimum (CBM) of one type valleys (e.g., either K or K′).

As the valley and spin degrees of freedom are interlocked,[13-15] the valley degeneracy can be lifted by using external magnetic field, magnetic doping or through contact with magnetic materials. The valley exchange splitting under an external magnetic field is typically very small as the Zeeman splitting is only ~0.11 meV/T.[16-18] Furthermore, it is not practical and energetically efficient to apply a strong magnetic field in devices. Therefore, recent theoretical research has focused on finding substrates and dopants that may produce large valley exchange splittings.

However, many large valley splittings predicted by theory have not been observed in experiment. One possible reason might be that the weak van der Waal (vdW) interlayer interaction is inadequate to grasp TMD layers on the substrate as tightly as assumed in theory, particularly with the existence of surface steps and kinks. The others possibility is the poor thermal stability of magnetization against thermal fluctuation which reduces the spin coherence. Naturally, it is better to implement magnetic sources in the TMD monolayers and it is hence important to seek for possible dopants that produce large valley splitting and have highly stable structural and magnetic properties. To this end, some recent experiments have shown the possibility of controlling dopants on high-quality samples.[19-22]

In this paper, we systematically investigate the structural and electronic properties of 3d metal adatoms on a defected tungsten disulfide monolayer ($WS_2$) by using the first-principles calculations and model simulations. While four of them have valley splitting larger than 100 meV, only $Mn/WS_2$ has a clean band gap and hence can be an excellent valley Hall system. $Mn/WS_2$ also has a large and tunable magnetic anisotropy energy to combat the thermal fluctuation and can be used at a high temperature. We further propose to use a hexagonal BN (hex-BN) overlayer to protect the novel magnetic and valley properties of $Mn/WS_2$. Our findings make a major step forward for the realization and utilization of the valley Hall effect in TMDs.

The density functional theory (DFT) calculations were carried out with the Vienna ab-initio simulation package (VASP) at the level of the spin-polarized generalized-gradient approximation (GGA).[28] A Hubbard U = 2.0 eV was added to take account of the correlation effect among d-electrons of 3d metals. The van der Waals correction was included for the description of dispersion forces. The interaction between valence electrons and ionic cores was considered within the framework of the projector augmented wave (PAW) method.[29,30] The energy cutoff for the plane wave basis expansion was set to 700 eV. To sample the two-dimensional Brillouin zone, we used a 7×7 k-grid mesh. All atoms were fully relaxed using the conjugated gradient method for the energy minimization until the force on each atom became smaller than 0.01 eV/Å.

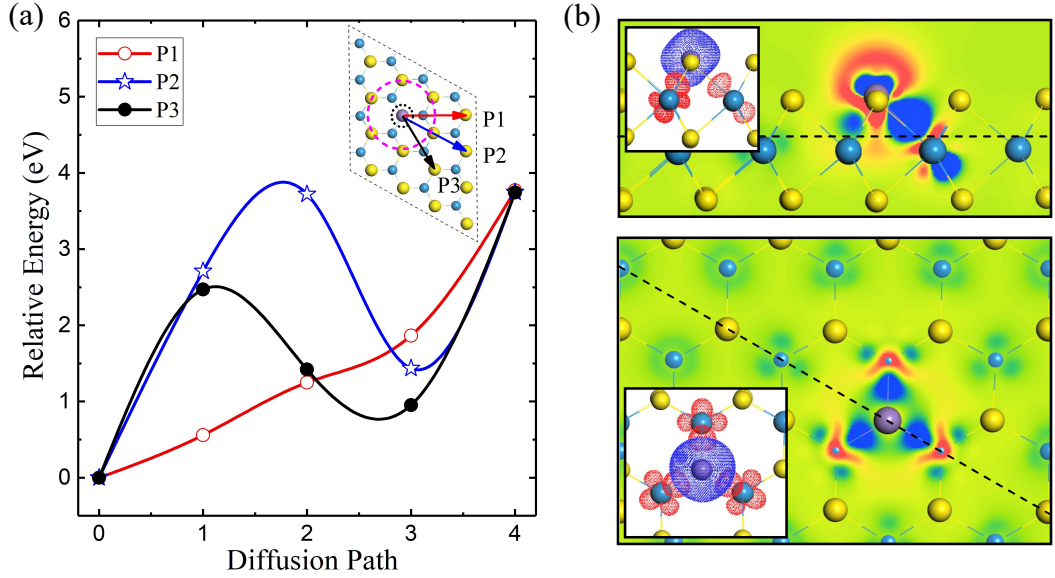

*Figure 1. (a) Three diffusion pathways for Mn atoms (inset) and the corresponding energy profiles. (b) Side and top views of charge density difference for Mn/WS$_2$, blue and red colors represent charge accumulation and depletion, respectively. (inset) Side and top views of spin density of Mn/WS$_2$, red and blue colors represent the minority and majority spin parts, respectively.*

As shown in Figure S1, the pristine WS$_2$ monolayer is a semiconductor with the CBM and VBM locating at the corners (K points) of the 2D hexagonal Brillouin zone[13,23] and a band gap of ~2eV. The strong SOC mostly originated from the 5d orbitals of W and the inversion symmetry breaking leads to noteworthy spin splitting, especially at the top of valence band near at K (K´), ~0.4 eV. The VBMs at opposite valleys have different spins due to the time-reversal symmetry, and the Berry curvatures around the two valleys have opposite signs, leading to an exact cancellation of the valley Hall currents and a vanishing anomalous Hall conductivity. To generate valley splitting, we use magnetic transition metal atoms (V, Cr, Mn, Fe, Co, Ni and Cu) which may break the time-reversal symmetry and lift the valley degeneracy.

As the segregation energy barrier for these adatoms on flat TMD monolayers are small, we adopt a 4×4 defected $WS_2$ monolayer as the template from which a sulfur atom is taken away. As known, sulfur cavities (Vs) may either naturally formed during the fabrication or deliberately produced in post processes.[24-26] To show the interaction between adatoms and the defected $WS_2$ monolayer, the energy cost of Mn diffusion and the charge density difference of Mn/$WS_2$ are shown in Figures 1a and 1b, respectively. One may see that Mn adatom takes the Vs site and stays almost in the same plane as the topmost sulfur atoms, i.e. the upper surface still appears to be reasonably flat. Electrons transfer out of Mn and its nearest W atoms to their intermediate regions, indicating the formation of strong chemical bonds between Mn and W atoms. Quantitatively, the binding energies of TM adatoms are calculated as:

$$\Delta E_{TM/WS_2} = E_{WS_2} + E_{TM} - E_{TM/WS_2}$$

where $E_{WS_2}$, $E_{TM}$ and $E_{TM/WS_2}$ are the total energies of defected $WS_2$, an isolated TM atom and TM/$WS_2$, respectively. As shown in Table I, values of $\Delta E_{TM/WS_2}$ are mostly larger than 2 eV except for Cr, indicating that TM adatoms are strongly anchored at the $V_S$ site. Meanwhile, three different pathways were considered for the diffusion of Mn atoms, as shown in the inset of Figure 1a. The energy barriers are overall large, 1.87 eV, 3.72 eV and 2.47 eV along P1, P2 and P3, respectively. It is conceivable that the adsorption geometry depicted in Figure 1 is rather stable against segregation.

*Table I. The binding energy (ΔE), total spin magnetic moments (Ms), magnetic anisotropy energy (MAE) and valley splitting (ΔKK′) of different TM adatoms on the defected WS$_2$ monolayer.*

|  | V | Cr | Mn | Fe | Co | Ni | Cu |
|---|---|---|---|---|---|---|---|
| ΔE | 3.63 | 1.93 | 2.94 | 3.44 | 3.85 | 4.26 | 2.44 |
| M$_S$/$\mu_B$ | 1.0 | 3.99 | 2.97 | 2.0 | 1.05 | 0.0 | 1.0 |
| MAE/meV | 2.41 | 0.25 | -0.50 | 1.04 | 2.66 | 0.0 | 2.71 |
| ΔKK′/ meV | 188 | -119 | 210 | 100 | 27 | 0.0 | 18 |

Now we check whether these 3d TM atoms can introduce strong valley splitting. Their electronic band structures are shown in Figure 2a and Figure S2. The presence of TM adatoms produces opposite effects on bands at the K and K′ valleys. Quantitatively, we define valley splitting as $\Delta KK′=E_K-E_{K′}$, where $E_K$ ($E_{K′}$) represents VBM at K (K′). As shown in Table I, (V, Cr, Mn, Fe)/WS$_2$ have huge valley splittings, especially for Mn/WS$_2$ which has the valley splitting as large as 210 meV, equivalent to the splitting induced by an external magnetic field of ~2000 T. Meanwhile, it has eminent advantage that the VBM is close to Fermi level for the realization of the valley spin Hall effect. Apparently, only Mn/WS$_2$ satisfies this requirement with a clean band gap, where all other systems have gap states from orbitals of adatoms. Furthermore, the key features of Mn/WS$_2$ retain in reasonably large ranges of Hubbard U (0-3eV, see Figure S3 in the Supplementary Information) and adsorption density (2-6.25%, see Figure S4 and the inset in Figure 2a). The robustness of the large valley splitting is also observed as we randomize the distribution of two Mn adatoms in a 7×7 supercell (see Figures S5 and S6 in the Supplementary Information). Hence, we believe that the giant valley splitting for Mn/WS$_2$ with sparse Mn atoms dopants can be realized. Below we take the case of 4×4 supercell with one Mn atom as example to further explore the corresponding electronic and valley properties of these systems.

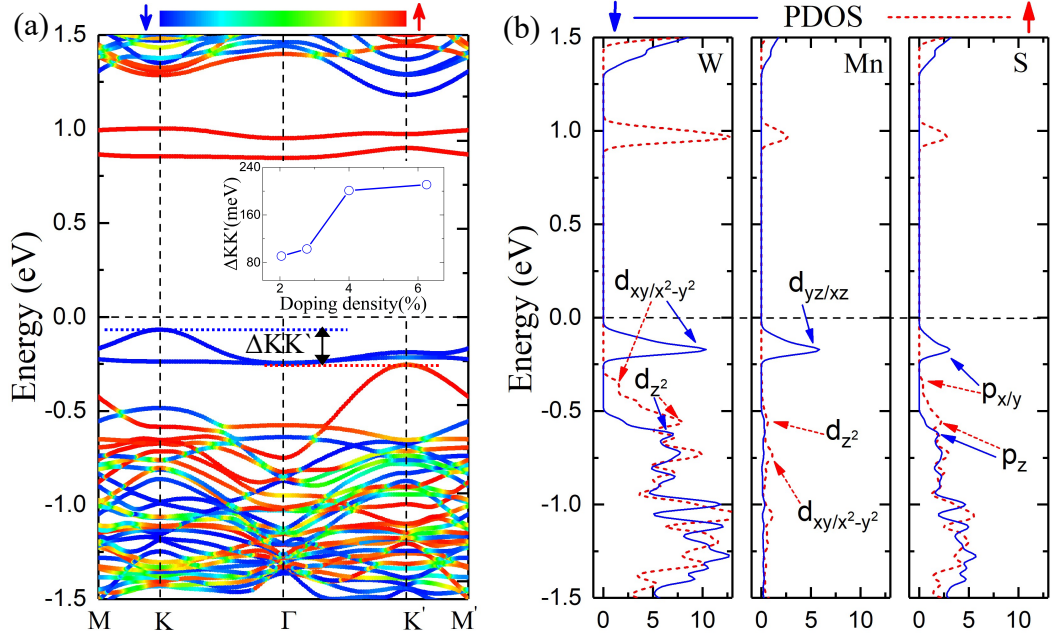

*Figure 2. (a) The band structure of Mn/WS$_2$ with SOC included. The red (blue) line represents the spin up (down) components. (The inset is the effect of dopants concentration on the valley splitting of Mn/WS$_2$. These four dopants concentration are corresponding to the case of one Mn atom periodic doping in 7×7, 6×6, 5×5 and 4×4 defected WS$_2$, respectively). (b) The corresponding projected density of states of all atoms, where dash-red lines and solid-blue lines are for the spin-up and spin-down channels, respectively.*

The curves of projected density of states (PDOS) of all atoms in Mn/WS$_2$ are shown in Figure 2b. One can see a strong magnetization locating at the Mn site, and a significant spin-polarization is also induced around the W atoms adjacent to Mn ($M_W \approx 0.3\ \mu_B$). This is also shown by the spin density in the insets of Figure 1b. The strong hybridization between Mn-$t_{2g}$ and W-$e_g$ orbitals imprint both large exchange and spin orbit coupling in the valence bands of WS$_2$. Even the $p_{x/y}$ orbitals of S atoms are strongly spin-polarized, as shown in Figure 2b, which are the essential causes for the valley splitting.

As a further step, we construct a simple Hamiltonian for WS$_2$ by fitting the DFT results as

$$H = H_o + H_{soc} + H_{ex} \qquad (1)$$

where $H_o$, $H_{soc}$ and $H_{ex}$ represent two band *k.p* gapped Dirac states, spin-orbit coupling and exchange interactions. Explicitly, these terms can be expressed in Pauli matrices as

$$H_o = \hbar v_f(\tau k_x \sigma_x + k_y \sigma_y) + \frac{\Delta_g}{2}\sigma_z \qquad (2)$$

$$H_{soc} = \tau s_z(\lambda_c \sigma_+ + \lambda_V \sigma_-) \qquad (3)$$

$$H_{ex} = -s_z \mu_B(B_z^c \sigma_+ + B_z^V \sigma_-) \qquad (4)$$

where $v_f$ is the Fermi velocity of the Dirac electrons, $\Delta_g$ is the gap, $k_x$ and $k_y$ are the electron wave vector, $\tau = 1(-1)$ for K (K′) valley, $s_z$ is the spin matrix along the z direction and $\lambda_c$ ($\lambda_V$) and $B_z^c$ ($B_z^V$) are the SOC parameters and exchange fields for the conduction (valence) band, respectively. The Pauli matrices, $\sigma_x = \begin{pmatrix} 0 & 1 \\ 1 & 0 \end{pmatrix}$, $\sigma_y = \begin{pmatrix} 0 & -i \\ i & 0 \end{pmatrix}$, $\sigma_z = \begin{pmatrix} 1 & 0 \\ 0 & -1 \end{pmatrix}$, $\sigma_+ = \begin{pmatrix} 1 & 0 \\ 0 & 0 \end{pmatrix}$ and $\sigma_- = \begin{pmatrix} 0 & 0 \\ 0 & 1 \end{pmatrix}$, represent the valley pseudospins. The corresponding fitting parameters of Mn ( V, Cr, Fe)/WS$_2$ are shown in Table S1. The strong SOC ($\lambda_V$) mostly comes from W atoms in the valence band, and the large negative effective exchange fields for conduction bands mean that an opposite spin polarization is induced near the TM atoms as shown in Fig 1b. Meanwhile, the large range of exchange fields for these four cases suggests that we can mix TM adatoms to tune the valley splitting.

With the giant valley splitting, we may selectively create valley polarization by shifting the Fermi level to slightly below the VBM. A transverse spin current arises as an in-plane electric field, E, is applied.[13,27] The anomalous Hall velocity of an electron in the *n*th band at a k-point depends on E and $\Omega_n(k)$ as $v \sim E \times \Omega_n(k)$, where $\Omega_n(k)$ is the Berry curvature of Bloch electrons determined according to $\Omega_n(k) = \hat{z} \cdot \nabla_k \times \langle u_n(k)|i\nabla_k|u_n(k)\rangle$ from the periodic part of Bloch wave functions, $|u_n(k)\rangle$. For Mn/WS$_2$, the distribution of the Berry curvature in Brillouin zone is shown in the left panel of Fig 3a. Obviously, $\Omega(k)$ ($=\sum_n f(\varepsilon_{n,k})\Omega_n(k)$, with $f(\varepsilon_{n,k})$ the Fermi-Dirac distribution function for the n[th] band at a k-point in the two-dimensional BZ) has opposite signs around the K and K´ points, typical for pristine and magnetized TMD monolayers. The large valley exchange splitting nevertheless allows us to use carriers only at K or K´ by shifting the Fermi level below VBM at the K-points with a bias. Indeed, the new distributions of Berry curvature with the Fermi level lying below the VBM at K but above the VBM at K´ only has the negative parts in the right panel of Figure 3a. The net Hall current can hence be produced in this circumstance.

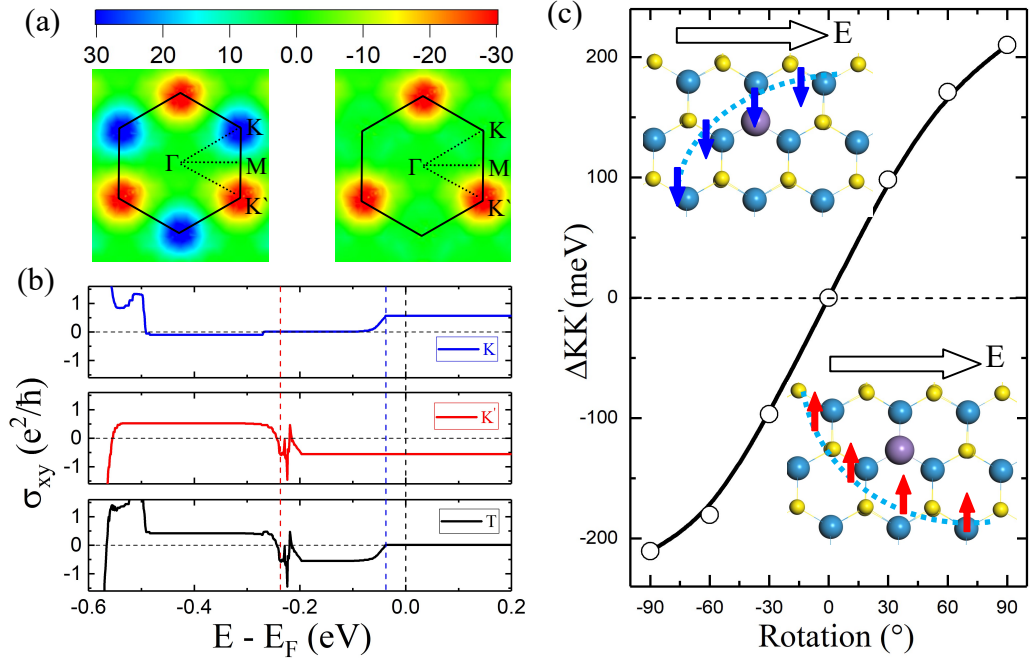

*Figure 3. (a) Left and right panels are the distributions of Berry curvature in the whole Brillouin zone of Mn/WS$_2$ with the Fermi level lying in the gap and intervalley, respectively. (b) The anomalous Hall conductivity $\sigma_{xy}$ as a function of Fermi level. (c) The variation of valley splitting ΔKK´ as the magnetization of Mn/WS$_2$ rotates. The insets show the schematics of the anomalous valley Hall effect (insets).*

Quantitatively, the Hall conductivity as a function of the position of the Fermi level is given by the integral of the Berry curvature of occupied states over the Brillouin zone.

$$\sigma_{xy} = \frac{e^2}{\hbar} \int_{BZ} \Omega(k) \frac{dk}{(2\pi)^2}$$

While $\sigma_{xy}$ is zero in a broad energy range as we move the Fermi level in the band gap due to the mutual cancellation between valleys, the presence of Mn induces a large valley splitting, i.e., a non-zero anomalous Hall conductivity can be observed right below the VBM at K. This is encouraging as one may use the degree of valley without the assistance of light. In addition, we also show the valley splitting as function of magnetization direction in Figure 3(c). One can see that the valley splitting can change continuously from 210 meV to -210 meV, following the rotation of magnetization direction. This is also useful for applications as one may reset the direction of magnetic moment with a pulse of magnetic field or a THz wave.

For the realization of valley anomalous Hall effect, it is important to have ferromagnetic ordering among magnetic sources and perpendicular magnetic anisotropy. Unfortunately, Mn/WS$_2$, has an in-plane easy axis, as indicated by its negative magnetic anisotropy energy (MAE= -0.50 meV), and a weak antiferromagnetic ordering among Mn adatoms in the ground state (c.f., Figure S5 in the supplementary information). However, the rigid band model analysis indicates that the MAE can be tuned to the positive side as soon as we move the Fermi level to slightly below the VBM, as shown in Figure 4(a). The curves in the lower panel indicate that crossing the VBM at the K-valley produces a positive blip of MAE, due to the removal of these spin-down (DD) states from the occupied region. This break the balance between contributions from different spin channels (UU, DD, UD and DU) and lead to a rapid change of MAE right below the Fermi level. The possibility of changing MAE in sign by moving down the Fermi level is actually in good coherence with the need of producing non-zero $\sigma_{xy}$ and it can be done by applying an external electric field (EEF, downward to surface as positive). Because of the weak screening between Mn and WS$_2$, adequate electron transfer between Mn and WS$_2$ can be easily achieved with a small field, as seen in the inset of Figure 4(b). Note that the calculated MAE curve closely follows the trend predicted by the rigid band model in Figure 4(a), and MAE increases rapidly to +6 meV (perpendicular easy axis) with a negative EEF ($\varepsilon \leq 0.5$ V/Å). Moreover, the same EEF also changes the weak antiferromagnetic coupling between Mn adatoms to the ferromagnetic regime, so all conditions for the realization of valley spin Hall effect are satisfied. It is plausible that the value of valley splitting and overall the band structure of Mn/WS$_2$ are barely affected by the EEF, as shown in Figure S7 in the Supplementary Information. Therefore, Mn/WS$_2$ is an excellent platform for the observation of valley Hall effect.

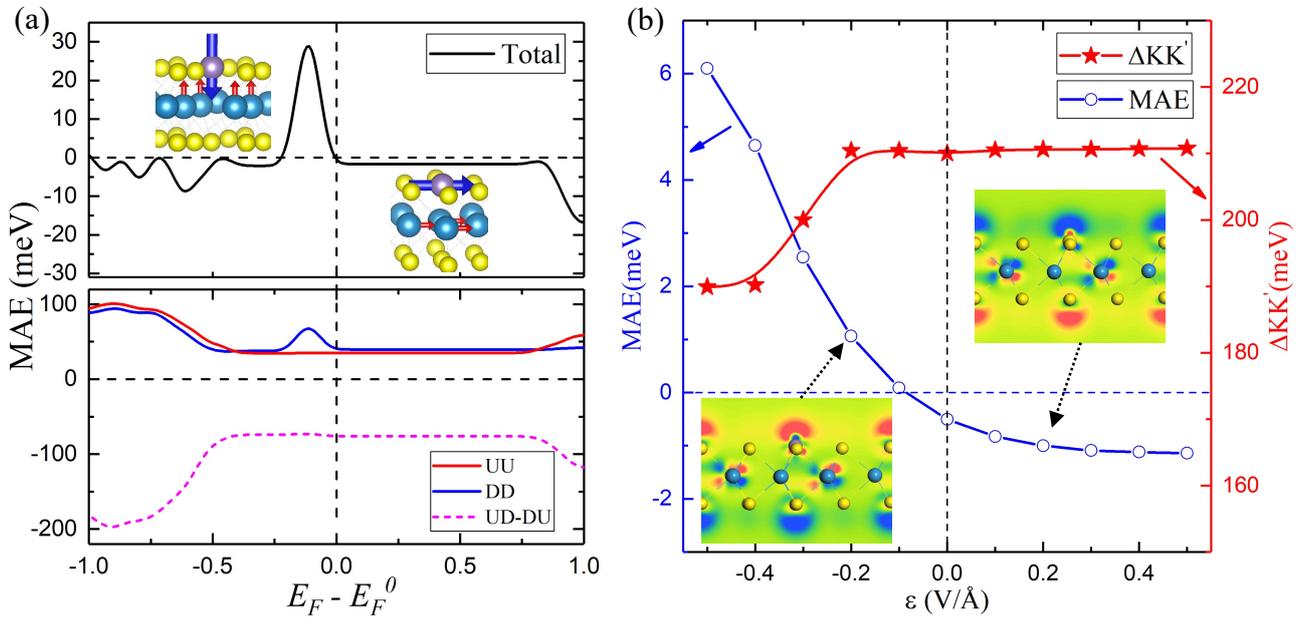

*Figure 4. (a) The Fermi level dependent total and spin channel decomposed MAEs of Mn/WS$_2$ from the rigid band model analyses. (b) The MAE and valley splitting of Mn/WS$_2$ as a function of external electric field ($\varepsilon$). The inset show the charge density difference of Mn/WS$_2$ caused by the external electric field of -0.2 and 0.2V/Å, and blue and red colors represent charge accumulation and depletion, respectively.*

Finally, we noted that Mn/WS$_2$ is chemically active and hence can be easily damaged by adsorbates such and H$_2$O and O$_2$ in the environment. Therefore, we propose to protect its valley features by putting a layer of hexagonal BN (hex-BN) on the top. Due to the van der Waals interaction, the Mn atoms are slightly pulled upward by 0.09 Å after hex-BN is added. Nevertheless, the hex-BN monolayer is a semiconductor with a band gap larger than 5 eV so its disturbance on the electronic property of Mn/WS$_2$ is negligible, as seen from the band structure of this van der Waals heterostructure in Figure S8 in the Supplementary Information. We further propose a conceptual device, as shown in Figure S9, to detect the valley anomalous Hall effect in experiments. The silicon substrate and hex-BN cover-layer provide a protective environment for Mn/WS$_2$ as well as channels for applying the bias. The Hall bar pattern with four Au electrodes allows the measurement of transvers voltage and longitudinal current.

In summary, we systematically investigated the possibility of achieving large valley splitting (> 100 meV) by using of 3d transition metal adatoms on the defected WS$_2$ monolayer. Mn/WS$_2$ is identified as a potential system for the realization of valley spin Hall effect without the assistance of light as it gives a valley splitting as large as 210 meV and, more importantly, has a clean band gap and appropriate band alignment with the Fermi level. Furthermore, we proposed to cover Mn/WS$_2$ with a hex-BN monolayer for the protection of its excellent valley and transport properties. Our studies suggest a realistic system for experimental verification and one practical avenue for exploring valley-spin physics.

**AUTHOR INFORMAYION**

**Corresponding Author**

* E-mail: wur@uci.edu.

**ACKNOWLEDGMENTS**

Work was supported by DOE-BES (Grant No. DE-FG02-05ER46237). Calculations were performed on parallel computers at NERSC.